\documentclass{book}

\usepackage{makeidx}
\usepackage{amsthm,amsmath,amssymb,amsfonts}
\usepackage{setspace,graphicx}
\usepackage{Generic}  
\usepackage[sort,longnamesfirst]{natbib}
\usepackage{url}

\numberwithin{equation}{chapter}
\theoremstyle{plain}

\usepackage{xcolor} 
\DeclareMathOperator*{\argmin}{arg\,min}

\newcommand{\vect}[1]{#1}

\usepackage{fancyvrb}

\newtheorem*{runningexample}{Running example}

\begin{document}

\chapter[Control Variates]{Control Variates for MCMC} \label{sec: CVforMCMC}
\begin{large}
\textbf{Leah South}\\
\textit{School of Mathematical Sciences, Queensland University of Technology, Australia}\\ \\
\textbf{Matthew Sutton}\\
\textit{School of Mathematics and Physics, University of Queensland, Australia}
\end{large}

\section{Motivation}

The focus of this chapter is on estimating an expectation $\mu = \mathbb{E}_{\pi}[f(x)] = \int_{\mathcal{X}} f(x) \pi(x) \mathrm{d} x$ where $\pi$ is a probability density on $\mathcal{X} \subseteq \mathbb{R}^d$ and $f:\mathcal{X} \rightarrow \mathbb{R}$ is a function of interest. This expectation could be a mean, variance, prediction or other quantity of interest. We are interested in estimating $\mu$ using output from a $\pi$-invariant Markov chain, $X^{(1)},X^{(2)},\ldots,X^{(n)}$. 

Recall that for stationary Markov chains, the Monte Carlo estimate $\hat{\mu}_{\text{MC}} = \frac{1}{n} \sum_{i=1}^n f(X^{(i)})$ follows the Markov chain central limit theorem so that for large $n$,
\begin{align*}
    \hat{\mu}_{\text{MC}} \approx \mathcal{N}\left(\mu, \frac{\sigma(f)^2}{n}\right),
\end{align*}
where $\sigma(f)^2$ is the asymptotic variance,
\begin{align}\label{eqn:asymptoticVar}
    \sigma(f)^2 = \text{var}\{f(X^{(i)})\} + 2\sum_{k=1}^{\infty}\text{cov}\{f(X^{(i)}),f(X^{(i+k)})\}.
\end{align}
For large enough $n$, the variance of $\hat{\mu}_{\text{MC}}$ depends only on two factors: the sample size $n$ and the asymptotic variance $\sigma(f)^2$. Using the dependence on $n$, one can reduce the variance of this estimator by a factor of $\tau$ by running the MCMC chain for $\tau$ times as many iterations. However, this can be prohibitively expensive when there is a high cost per iteration of the MCMC algorithm or per evaluation of $f$. 

An alternative way to improve the precision of $\hat{\mu}_{\text{MC}}$ is to reduce the asymptotic variance $\sigma(f)^2$. Chapters 1 and 2 describe ways in which redesigning the sampler can reduce autocorrelation in the chain and therefore reduce $\sigma(f)^2$. This chapter will instead reduce the variance by adjusting $f$ in such a way that the expectation remains the same but the variance of the estimator is reduced, using so-called control variates. 

This chapter aims to provide a tutorial on the use of control variates for post-processing of MCMC. We begin by describing the key steps in building a successful control variate in Section \ref{sec:CVBasics}. Next, we describe three strategies that have emerged for general construction. These strategies are well-suited to the following contexts:
\begin{itemize}
    \item Where there is a $\pi$-invariant Markov chain with tractable next-step-ahead conditional expectation $\mathbb{E}[h(X^{(i+1)})|X^{(i)}=x]$ for some $h: \mathcal{X} \rightarrow \mathbb{R}$ (Section \ref{sec:PoissonCV}).
    \item Where $\log \pi$ is continuously differentiable and, ideally, gradient-based methods have been used for the MCMC sampling (Section \ref{sec:GBCV}).
    \item Where Metropolis--Hastings (MH) MCMC has been used for sampling (Section \ref{sec:MH}).
\end{itemize}
We conclude the chapter with further practical advice and a realistic example in Section \ref{sec:practical}.

\section{Control variate basics}\label{sec:CVBasics}

In control variates, the vanilla Monte Carlo estimator is replaced with
\begin{align}\label{eqn:CV}
\hat{\mu}_{\text{CV}} = \frac{1}{n} \sum_{i=1}^n \left[f(X^{(i)}) - u(X^{(i)})\right] + \int_{\mathcal{X}} u(x) \pi(x) \mathrm{d} x,
\end{align}
where $u: \mathcal{X} \rightarrow \mathbb{R}$ is a function with known expectation under $\pi$. For large $n$, $\hat{\mu}_{\text{CV}} \approx \mathcal{N}(\mu,\frac{\sigma(f-u)^2}{n})$ by the Markov chain central limit theorem, so the goal is to optimise $u$ to achieve maximal variance reduction.

\subsection{Finding functions with known expectation}

The first challenge in finding a suitable control variate is in finding candidate functions with known expectation under $\pi$. This has historically been a barrier to implementing control variates in MCMC and is likely part of the reason why control variates have not been used more widely in the MCMC community since their early introduction \citep[see e.g.][]{Green1992,Philippe2001}.

Fortunately, several broadly applicable approaches to finding functions with known expectation under $\pi$ have recently been developed to overcome this hurdle. This will be the focus of Sections \ref{sec:PoissonCV}-\ref{sec:MH}. The aim of these sections is to describe three general frameworks for deriving families of control variates that can be used for post-processing, rather than providing an exhaustive list of all control variates that have been proposed for MCMC.

\subsection{Optimisation}\label{ssec:optimisation}
The control variates introduced in Sections \ref{sec:PoissonCV}-\ref{sec:MH} all depend on a set of parameters. For most control variates, the parameters come in the form of additive constants and multiplicative factors. Multiplicative factors are so important that one will often see the control variate equation \eqref{eqn:CV} written as $\hat{\mu}_{\text{CV}} = \frac{1}{n} \sum_{i=1}^n \left[f(X^{(i)}) - \theta u(X^{(i)})\right] + \theta \int_{\mathcal{X}} u(x) \pi(x) \mathrm{d} x$ where $\theta$ is a multiplicative factor to be optimised. We choose to keep this dependence on parameters less explicit to allow for more sophisticated control variates. However, we would like to remind readers that parameters are important for maximising the potential for variance reduction and reducing the risk of increasing the variance. If you find multiple control variates $u_j(x)$ for $j=1,\ldots,J$, each with zero expectation under $\pi$ without loss of generality, then you can combine them using $u(x) = \alpha + \theta_1 u_1(x) + \cdots + \theta_J u_J(x)$ so that $\mathbb{E}_{\pi}[u(x)] = \alpha$, where $\alpha \in \mathbb{R}$ and $\theta \in \mathbb{R}^J$. The challenge is in optimising these parameters. For full generality, we write this as selecting $u \in \mathcal{U}$.

Ideally we would like to select $u$ to minimise the expected mean squared error (MSE) of the estimator, i.e.\ $\mathbb{E}\left[(\hat{\mu}_{\text{CV}} - \mu)^2\right]$, where the expectation here is with respect to the Markov chain. Evaluating the expected MSE is a more challenging problem than the original problem of estimating $\mu$, so we instead optimise approximations to the MSE or asymptotic variance. Below we describe some of the most popular approaches\footnote{All approaches described in this chapter are based on using the same samples for optimising $u \in \mathcal{U}$ and evaluating the estimator \eqref{eqn:CV}. This can technically introduce bias for otherwise unbiased estimators, such as those based on unbiased MCMC \citep[Chapter 17;][]{Jacob2020}, but typically leads to lower MSE. The bias is most significant when the optimisation is complex, for example with many parameters, and the sample size $n$ is small. It is important to use independent samples for optimisation and evaluation if applying the estimators to unbiased MCMC.}.

One approach designed for independent samples is to minimise the empirical variance \citep{Belomestny2022}, i.e.\ to select $u$ according to the minimisation problem
\begin{align}
\argmin_{\substack{u \in \mathcal{U}}} \frac{1}{n-1}\sum_{i=1}^n \left[f(X^{(i)}) - u(X^{(i)}) - \left(\frac{1}{n}\sum_{j=1}^n f(X^{(j)}) - u(X^{(j)})\right)\right]^2.
\end{align}
Alternatively, one could perform least squares which is equivalent to minimising an upper bound on the empirical variance. The optimisation in this case becomes
\begin{align}
\argmin_{\substack{u \in \mathcal{U}}} \sum_{i=1}^n \left(f(X^{(i)}) - u(X^{(i)})\right)^2.
\end{align}
This is a popular and typically fast approach that can outperform empirical variance minimisation \citep{Si2020}. If $u(x) = \alpha + \theta_1 u_1(x) + \cdots + \theta_J u_J(x)$ then this amounts to performing ordinary least squares on the multiple linear regression problem with $f$ as the response and $u_j(x)$ for $j=1,\ldots,J$ as the predictors, so there is a closed-form solution. This also gives some intuition into when control variates will perform well. Control variates typically do not perform well when $f$ is not well predicted by $u_1, \ldots, u_J$, for example when $f$ is an indicator function. Empirical variance minimisation and least squares are well-suited to independent and identically distributed samples. They can also perform well in the context of MCMC but they do not directly attempt to minimise the asymptotic variance.

A more suitable approach in the context of MCMC would be to minimise the asymptotic variance $\sigma(f-u)^2$. Since this is not tractable, approaches based on minimisation of a finite-sample estimate of the asymptotic variance can be used (see, e.g., Chapter 5 or \citet{Flegal2010}). Minimisation of spectral variance estimates \citep{Belomestny2020,Brosse2019} and batch means estimates \citep[see e.g.][]{Hammer2008} have been considered. 

In an entirely different direction, some non-parametric control variates are optimised by selecting $u$ to be a minimum-norm interpolant to $f$ \citep{Oates2017}, so the optimisation is decoupled from any measure of asymptotic variance for the sampler.

One can introduce regularisation into any of these approaches. This can be particularly helpful when the number of parameters to be optimised is high relative to the effective sample size of the chain. We have found the use of an $\ell_1$ penalty to be particularly helpful when the control variates are of the form $u(x) = \alpha + \theta_1 u_1(x) + \cdots + \theta_J u_J(x)$ and $J\gg 1$ \citep{South2019}.

The optimisation step is important. Poorly optimised $u$ can lead to increased variance relative to the vanilla Monte Carlo estimator and well-optimised $u$ can help to maximise the variance reductions achieved. For example, consider the simple control variates $u(x) = \alpha + \theta u_1(x)$ where $\mathbb{E}_{\pi}[u_1(x)] = 0$. If $f$ and $u_1$ have correlation one under $\pi$ then selecting the optimal $\alpha$ and $\theta$ would lead to zero-variance (exact estimators) and the estimator $\hat{\mu}_{\text{CV}}$ would return $\mu$ for finite $n$. This can be thought of as selecting $u=f$ when $f \in \mathcal{U}$, which is achieved by several of the optimisation approaches described above (with regularised optimisation being an exception). On the opposite end of the spectrum, we would hope to estimate $\theta \approx 0$ if $f$ and $u_1$ are uncorrelated. 

\subsection{A running example}\label{ssec:example}

We use a bivariate Gaussian example throughout this chapter to demonstrate the use of control variates. Our goal will be to estimate $\mathbb{E}_{\pi}[x_1]$, for which the true expectation is $0$, 
when ${x} \sim \mathcal{N}\left(\begin{pmatrix} 0 \\ 0\end{pmatrix}, \Sigma\right)$ and $\Sigma = \begin{bmatrix}1 & \rho \tau\\ \rho\tau & \tau^2\end{bmatrix}$. Following \citet{Dellaportas2012}, we choose $\rho = 0.99$ and $\tau^2 = 10$. For our MCMC sampler, we use $n=100$ iterations of a random-walk MH algorithm with covariance $\Sigma$ and starting point $X^{(1)} = (0.5,0.5)^\top$.

We choose this example because we can apply all of the control variates presented in this chapter to the problem and we know the true expected value for comparative purposes. We do not intend for readers to draw conclusions on which control variates are best based on this example alone. In fact, this application is perfectly suited to the methods in \ref{ssec:ZVCV}, \ref{ssec:SECF} and certain variations of methods in \ref{sec:PoissonCV}. We use this example for pedagogical reasons and a more realistic example is presented in Section \ref{ssec:BVS}.

Below we provide R code for estimating the expected value $\mathbb{E}_{\pi}[x_1]$ using vanilla Monte Carlo and two different control variates. 
Full R code for this example can be found at \url{https://github.com/LeahPrice/CVBookChapter/}. The code below simply refers to the matrix \verb|SigmaInv| ($\Sigma^{-1}$) and the output \verb|X| which is the $n \times 2$ matrix containing the samples $X^{(1)},X^{(2)},\ldots,X^{(n)}$ from the Markov chain.

\begin{Verbatim}[numbers=left,xleftmargin=5mm]
> # Evaluating the integrand and performing vanilla Monte Carlo
> f <- X[,1]
> mean(f)
[1] -0.03997147
> 
> # Getting the score function
> grads <- -X%*%SigmaInv
> 
> # Control variates with least squares
> coef <- lm(f ~ grads)$coef
> mean(f - grads%*%coef[-1])
[1] 2.9865e-16
> 
> # Control variates with batch means
> coef <- optim(c(0,0),function(theta) bm(f-grads%*%theta)$se)$par
> mean(f - grads%*%coef)
[1] -9.396387e-10
\end{Verbatim}

The control variate used in this example is the score function $\nabla_x\log\pi(x)$. Like many other methods in computational statistics \citep[see][for example]{Fan2006}, we are exploiting the property that the score function has zero expectation, i.e.\ that $\mathbb{E}_{\pi}[\nabla_x\log\pi(x)]=0$, where in this case $\nabla_{x}\log\pi(x) = -\Sigma^{-1}x$. This leads us to control variates of the form $u(x) = \alpha + \theta_1 \frac{\partial}{\partial x_1}\log\pi(x) + \theta_2 \frac{\partial}{\partial x_2}\log\pi(x)$ and therefore estimators of the form $\hat{\mu}_{\text{CV}} = \frac{1}{n}\sum_{i=1}^n f(X^{(i)}) - \theta_1 \frac{\partial}{\partial x_1}\log\pi(X^{(i)}) - \theta_2 \frac{\partial}{\partial x_2}\log\pi(X^{(i)})$.
We will later see that this family of control variates exactly matches the first-order zero-variance control variates of Section \ref{ssec:ZVCV}.

Calculation of the score function is done in post-processing for this example, on line 7. This calculation is straightforward for a Gaussian target but this would be a bad idea for more complicated problems with expensive gradients. In those cases, we would ideally be using gradient-based sampling and storing the gradients alongside the chain.

The next step is to perform the optimisation. There is substantial flexibility here but we will consider two of the approaches described in Section \ref{ssec:optimisation}: least squares and minimisation of the estimated asymptotic variance. We use least squares to estimate $\alpha$ and $\theta$ through the \verb|lm| function on line 10. The calculation of the corresponding estimator is then performed on line 11. On line 15, we perform an alternative optimisation based on minimising the batch means estimate of the standard error using the R package batchmeans \citep{R_batchmeans}. The intercept $\alpha$ is irrelevant in this case, so we are simply estimating $\theta$. We use R's base \verb|optim| function starting with $\theta=(0,0)^\top$, which is equivalent to starting with no control variates.

Looking at the estimators for the two control variates on lines 12 and 17, we can see that the estimates using control variates are incredibly precise relative to the small number ($n=100$) of samples used.

\section{Control variates using the Poisson equation} \label{sec:PoissonCV}

We will now consider our first approach to generating control variates that is broadly applicable to MCMC. Here we will use control variates of the form $u = \alpha + h - Ph$, where $h:\mathcal{X}\rightarrow \mathbb{R}$ is a user-specified function and $P$ is the one-step-ahead conditional expectation operator $Ph(x) = \mathbb{E}[h(X^{(i+1)})|X^{(i)} = x]$ of a $\pi$-invariant Markov chain. This is a valid control variate because the $\pi$-invariance of $P$ signifies that $\mathbb{E}_{\pi}[h(x) - Ph(x)] = 0$. Existing literature assumes $P$ is the one-step-ahead conditional expectation for the same Markov chain that was used for sampling, but we note that this is not strictly necessary. Evaluation of $Ph$ can be challenging so we expand on this in Section \ref{ssec:EvalPoissonCV}. 

Theoretically, the optimal function $h$ that leads to perfect (zero-variance) estimators could be obtained by solving Poisson's equation. For discrete-time Markov chains, Poisson's equation is
\begin{align}\label{eqn:PoissonEq}
    Ph(x) - h(x)= -f(x) + \mathbb{E}_{\pi}[f(x)],
\end{align}
where $h$ is referred to as the solution to the Poisson equation for $f$
or as the ``fishy function" because the French word Poisson translates to fish \citep{Douc2022}. The fishy function is typically intractable so Section \ref{ssec:ApproximatingFishy} describes approaches to approximating it.

The approach of selecting control variates based on $h - Ph$ and approximately solving Poisson's equation to choose $h$ was first considered by \citet{Henderson1997} in the context of continuous-time Markov processes. The idea has since been extended to discrete-time Markov chains.

\subsection{Evaluating the control variate}\label{ssec:EvalPoissonCV}

Let us first describe some examples where $h - Ph$ is tractable. One example is finite state-spaces, where $P$ is the Markov transition matrix. Another example, considered in \citet{Dellaportas2012}, is conditionally conjugate random-scan Gibbs sampling. This is demonstrated for the Gaussian example from \citet{Dellaportas2012} below.

\begin{runningexample}

We will now develop control variates of the form $h - Ph$ based on an alternative Markov chain for our Gaussian example. Specifically, we know that a random-scan Gibbs sampler that randomly updates either $x_2 | x_1 \sim \mathcal{N}(\rho \tau x_1, \tau^2 (1-\rho^2))$ or $x_1 | x_2 \sim \mathcal{N}(\rho x_2/\tau,1-\rho^2)$ is $\pi$-invariant. The operator $P$ for this alternate Markov chain is tractable for many possible choices of $h$. We will consider $h(x) = \theta_1 x_1 + \theta_2 x_2$ as in \citet{Dellaportas2012}. To make our calculations easier, we will separately calculate $Px_1$ and $Px_2$ first. Starting with $Px_1$, we have
\begin{align*}
    P x_1 &= 0.5 x_1 + 0.5\mathbb{E}_{x_1 \sim \mathcal{N}(\rho x_2/\tau,1-\rho^2)}[x_1]\\
    &= 0.5 x_1 + 0.5\rho x_2/\tau.
\end{align*}
Now for $Px_2$, we have
\begin{align*}
P x_2 &= 0.5 x_2 + 0.5\mathbb{E}_{x_2 \sim \mathcal{N}(\rho \tau x_1, \tau^2 (1-\rho^2))}[x_2]\\
    &= 0.5 x_2 + 0.5\rho \tau x_1.
\end{align*}
Putting this together, we have control variates of the form $u(x) = \alpha + h(x) - Ph(x) = \alpha + 0.5 \theta_1(x_1 - \rho x_2/\tau) + 0.5 \theta_2 (x_2 - \rho \tau x_1)$. We will proceed with this choice of control variate for our simple problem of estimating the mean of a Gaussian distribution, but we could extend this to higher-order polynomial $h$ if we had a more complex problem.

Section \ref{ssec:ApproximatingFishy} will describe some optimal approaches for selecting $\theta$ based on the situation where we are using the same $P$ that was used for sampling and the Markov chain is reversible. Since we are not in that situation here, we will instead resort to more standard estimators for the coefficients, specificaly least squares for simplicity. Code for this is shown below.

\begin{Verbatim}[numbers=left,xleftmargin=5mm]
> CVs <- 0.5*cbind(X[,1] - rho*X[,2]/tau,X[,2] - rho*tau*X[,1])
> coef <- lm(f ~ CVs)$coef
> mean(f - CVs%*%coef[-1])
[1] -6.77236e-17
\end{Verbatim}

\end{runningexample}

Unfortunately $P$ is not analytically tractable for many popular MCMC samplers, including most MH algorithms. When $P$ is not tractable, it could theoretically be replaced with an unbiased estimate. Let us consider this now in the context of MH-MCMC. Following \citet{Tsourti2012}, for MH-MCMC we have
\begin{align*}
    h(x) - Ph(x)&= h(x) - \int_{\mathcal{X}} \left[\alpha(x,y) h(y) + (1-\alpha(x,y)) h(x)\right] q(y|x) \mathrm{d}y\\
     &= h(x) - \int_{\mathcal{X}} \alpha(x,y) h(y)q(y|x) \mathrm{d}y - h(x) \int_{\mathcal{X}} (1-\alpha(x,y)) q(y|x) \mathrm{d}y\\
     &= \int_{\mathcal{X}} \alpha(x,y) (h(x) - h(y)) q(y|x) \mathrm{d}y,
\end{align*}
where $\alpha(x,y) = \text{min}(1,R(x,y))$ is the acceptance probability, $R(x,y) = \frac{\pi(y)q(x|y)}{\pi(x)q(y|x)}$ is the MH-ratio and $q(y|x)$ is the density of the proposal distribution evaluated at $y$ given current value $x$. This is not analytically tractable but it is feasible to get an unbiased estimate by storing and using the proposals and acceptance probabilities from each step in the chain \citep{Alexopoulos2023}. If $X^{(i+1)^*} \sim q(\cdot|X^{(i)})$ denotes the proposed value moving from $X^{(i)}$ to $X^{(i+1)}$, then the estimator could be written as 
\begin{align}\label{eqn:Alexo}
\hat{\mu}_{CV} = \frac{1}{n-1}\sum_{i=1}^{n-1} f(X^{(i)}) - \theta \alpha(X^{(i)},X^{(i+1)^*})[h(X^{(i)}) - h(X^{(i+1)^*})],
\end{align}
where $\theta$ could be optimised using any approach from Section \ref{ssec:optimisation}. This approach is similar to the control variates for MH samplers in Section \ref{sec:MH}.

High variance in unbiased estimates for $u(x)$ could negatively affect the performance of the control variates and unfortunately advice in this direction is limited in the literature. To keep the variance of the aforementioned control variate down, \citet{Alexopoulos2023} introduce an additional control variate and \citet{Tsourti2012} use multiple importance sampling. In the absence of low-variance unbiased estimators for $h - Ph$, it may be better to stick with control variates based on analytically tractable $P$.

As mentioned earlier, we can use the operator $P$ for any $\pi$-invariant Markov chain that need not match the one used for sampling. However, it would not be a good idea to use control variates based on \eqref{eqn:Alexo} if MH MCMC had not been used for sampling or if the proposals and acceptance probabilities had not been stored. Doing so would require an additional $n$ draws from $q$ and an additional $n$ or up to $2n$ evaluations of $\pi$ (up to a normalising constant), depending on whether these calculations were stored for $X^{(i)}$ with $i=1,\ldots,n$. We do not want our control variate to require additional evaluations of (unnormalised) $\pi$.

\subsection{Approximating the fishy function}\label{ssec:ApproximatingFishy}

Recall that selecting $h$ as the solution to the Poisson equation for $f$ would lead to perfect, zero-variance estimators. This is not generally feasible because the Poisson equation can only be solved exactly in simple cases such as when $\mathcal{X}$ is a small discrete state space, making the Poisson equation a tractable linear system. Fortunately, a variety of techniques can be used to approximate the solution. We do not need a consistent estimator to the fishy function for use in control variates. We simply need one that is similar enough to the true function that it can lead to a reduced-variance estimator.

We can introduce any number of approximations into the Poisson equation when solving it for $h$, including replacing $f$ with an approximation $\tilde{f}$, replacing $\pi$ with an approximation $\tilde{\pi}$ and using a next-step-ahead conditional expectation $\tilde{P}$ that is $\tilde{\pi}$-invariant (rather than $\pi$-invariant). This approach was first employed to develop control variates for Markov processes in \citet{Henderson1997}, where they solved the Poisson equation exactly for a simple Markov process targeting an approximation $\tilde{\pi}$. 
More recently, \citet{Mijatovic2018} partitioned a continuous state space $\mathcal{X}$ into $m+1$ measurable sets and solved the Poisson equation exactly for a Markov chain on this partitioned space. A challenge with this approach is that approximation of $P$ for the partitioned space requires Monte Carlo methods with further evaluations of $\pi$ (up to a normalising constant).
\citet{Alexopoulos2023} developed an approach for estimating marginal means from certain MH samplers using numerical integration to approximately solve Poisson's equation for a simpler problem based on a Gaussian approximation to $\pi$.

Rather than trying to derive the complete functional form of $h$ from Poisson's equation, \citet{Dellaportas2012} propose to consider approximations $h(x) = \sum_{j=1}^k \theta_j h_j(x)$ for the fishy function. 
When the goal is to estimate marginal means, e.g.\ $f(x) = x_1$, the authors propose to use $h_j = x_j$ for all $j \in \{1,\ldots,d\}$ for which $Px_j$ or a reasonable unbiased estimator is available. This can be extended to higher-order terms when estimating high-order moments, as long as $Ph(x)$ is tractable for those terms. 

\citet{Dellaportas2012} propose a consistent estimator for the optimal coefficient $\theta$ for the case where the Markov chain is reversible, based on the relationship between the fishy function and expressions for the asymptotic variance. This approach to optimising $\theta$ is only sensible when $P$ is from the Markov chain actually used for sampling. Their estimator for $\theta$ is
\begin{align}\label{eqn:DellaTheta}
    \hat{\theta} = \hat{K}^{-1} \left[\left\{\frac{1}{n} \sum_{t=1}^n f(X^{(t)})({H}(X^{(t)}) + P{H}(X^{(t)}))\right\} 
- \left\{\frac{1}{n} \sum_{t=1}^n f(X^{(t)})\right\}\left\{\frac{1}{n} \sum_{t=1}^n {H}(X^{(t)}) + P{H}(X^{(t)})\right\}\right],
\end{align}
where ${H}(x) = (h_1(x),h_2(x),\ldots,h_k(x))^\top$ and
\begin{align}
    \hat{K}_{ij} = \frac{1}{n-1}\sum_{t=2}^n \{h_i(X^{(t)}) - Ph_i(X^{(t-1)})\}\{h_j(X^{(t)}) - Ph_j(X^{(t-1)})\}.
\end{align}
This makes the final estimator 
\begin{align}
    \hat{\mu}_\text{CV}= \frac{1}{n}\sum_{t=1}^n f(X^{(t)}) -  [{H}(X^{(t)}) - P{H}(X^{(t)})]\hat{\theta}.
\end{align}
Their estimator has been shown to have lower asymptotic variance than estimators based on batch means for reversible Markov chains \citep[equation 35 of][]{Dellaportas2012}, with the difference being largest in cases with high autocorrelation or when $|f|$ is large. This is also demonstrated empirically in their paper and in the extended note associated with the paper \citep{Kontoyiannis2009}.

\citet{Dellaportas2012} implicitly set $\alpha=0$ but allowing this to vary can lead to semi-exact estimators. For example, when $P$ is the next-step-ahead conditional expectation from random-scan Gibbs sampling and $\pi$ is Gaussian, using least squares to estimate $\alpha$ and $\theta$ for first-order polynomial $h(x)$ leads to exact estimators for first-order polynomial $f$, regardless of the Markov chain used to obtain the samples.

\subsection{Using $P$ from a Markov process}

Not needing to use the same $P$ that was used for sampling opens up the possibility of using $P$ based on an alternative $\pi$-invariant Markov chain or even a $\pi$-invariant Markov process. For any sufficiently regular $\pi$-invariant Markov process with infinitesimal generator $\mathcal{A}$ such that $\mathcal{A}h(x) = \lim_{t\rightarrow 0} \frac{\mathbb{E}[h(X^{(t)})|X^{(0)} = x] - h(x)}{t}$, we know $\mathbb{E}_{\pi}[\mathcal{A}h(x)] = 0$ under mild regularity conditions on $h$. One could therefore use control variates $u = \alpha + \mathcal{A}h(x)$. 

Estimators would be exact if we could solve for $h$ in Poisson's equation for continuous-time Markov processes,
\begin{align*}
    \mathcal{A}h(x) = -f(x) + \mathbb{E}_{\pi}[f(x)].
\end{align*}
As with discrete-time Markov chains, this is not tractable in practice but approximations have been considered for the purpose of control variates \citep{Mira2003}.

The Stein-based control variates that we will consider in Section \ref{sec:GBCV} can be thought of as using the $\mathcal{A}$ from an alternative Markov process, specifically overdamped Langevin diffusion. The optimisation of $h$ is typically based on optimisation of some criterion as per Section \ref{ssec:optimisation}, rather than directly attempting to solve the Poisson equation.

\section{Stein-based control variates}\label{sec:GBCV}

This section describes control variates that arise through the use of Stein operators. All of the control variates described in detail here use the score function, that is $\nabla_x \log \pi(x)$. These methods are a natural choice when gradient-based sampling has been performed, as in Chapter 2. 
Theoretically these control variates can be applied to estimate the expectation of any function $f$, but their performance is typically poor when $f$ is not smooth. This makes these methods good candidates for improving estimates of posterior means but not posterior probabilities, for example.

The main idea behind the methods in this section is to use a Stein operator $\mathcal{L}$ which gives functions with zero expectation under $\pi$. Specifically, consider control variates of the form 
\begin{align} \label{eqn:SteinCV}
    u(x) = \alpha + \mathcal{L} h(x),
\end{align}
where $\alpha \in \mathbb{R}$ is a constant, $h(x)$ is a function to be specified and $\mathcal{L}$ is a so-called Stein operator which signifies that $\mathbb{E}_{\pi}[\mathcal{L} h(x)] = 0$ and $\mathbb{E}_{\pi}[u(x)] = \alpha$ under regularity conditions elaborated on later in the chapter. For clarity, the resulting estimator combining this choice of control variates \eqref{eqn:SteinCV} with the control variate estimator \eqref{eqn:CV} is $\hat{\mu}_{\text{CV}} = \frac{1}{n} \sum_{i=1}^n \left[f(\theta_i) - \mathcal{L} h(x_i)\right]$. This framework was formally introduced by \citet{Oates2017} and is a generalisation of the zero-variance control variates (ZVCV) method that originally appeared in the physics literature \citep{Assaraf1999} and was introduced with further theory for the statistics community by \citet{Mira2013}. 

There is flexibility in the choice of Stein operator, with some options described in Section 2.2 of \citet{Anastasiou2023} and a gradient-free alternative presented in \citet{Han2018}. Popular choices are the first and second-order Langevin Stein operators, denoted here as $\mathcal{L}^{(1)}$ and $\mathcal{L}^{(2)}$ respectively, which are defined as
\begin{align} \label{eqn:LangevinStein1}
\begin{split}
    \mathcal{L}^{(1)} h(x) &= \nabla_x \cdot h(x) + h(x) \cdot \nabla_x \log \pi(x),
\end{split}
\end{align}
for $h(x):\mathcal{X}\rightarrow \mathbb{R}^d$ and 
\begin{align} \label{eqn:LangevinStein2}
\begin{split}
    \mathcal{L}^{(2)} h(x) &= \Delta_x h(x) + \nabla_x h(x) \cdot \nabla_x \log \pi(x),
\end{split}
\end{align}
for $h(x):\mathcal{X}\rightarrow \mathbb{R}$. We highlight that the function $h(x)$ must be vector-valued for \eqref{eqn:LangevinStein1} and real-valued for  \eqref{eqn:LangevinStein2}. The problem of selecting $u \in \mathcal{U}$ has now become a challenge of selecting $\alpha \in \mathbb{R}$ and $h \in \mathcal{H}$. The notation $\Delta$ represents the Laplacian operator $\nabla_x \cdot \nabla_x$, which is the sum of second-order gradients with respect to $x$.

These control variates are sometimes referred to as gradient-based control variates because they use $\nabla_x \log \pi(x)$. These gradients are already available when gradient-based sampling has been performed and they can be applied in Bayesian inference where the normalising constant of $\pi$ is unknown. The methods can also be applied when unbiased estimates of the gradients are available \citep[see][for example]{Baker2019,Hodgkinson2020}, although less is understood about their performance in this context.

These are valid control variates, in the sense that they have known expectation, under mild regularity conditions. We will focus here on $\mathcal{X} = \mathbb{R}^d$. First-order Langevin Stein operators require that $\int_{\mathcal{X}} |h(x)| \pi(x) \mathrm{d} x < \infty$ and $\int_{\mathcal{X}} |\mathcal{L}^{(1)}h(x)| \pi(x) \mathrm{d} x < \infty$, while second-order Langevin Stein operators require that $\int_{\mathcal{X}} |\nabla_x h(x)| \pi(x) \mathrm{d} x < \infty$ and $\int_{\mathcal{X}} |\mathcal{L}^{(2)}h(x)| \pi(x) \mathrm{d} x < \infty$ \citep[Proposition 3,][]{Oates2022}. 
For bounded domains, refer to Assumption 2 of \citet{Oates2017} for first-order Stein operators and Equation 9 of \citet{Mira2013} for second-order Stein operators.

The main differences between existing gradient-based control variates are in how the function class $\mathcal{H}$ is specified. It is the class of $Q$th order polynomials in ZVCV \citep{Assaraf1999,Mira2013}, a reproducing kernel Hilbert space in control functionals \citep[CF,][]{Oates2017}, and a combination of these parameteric and non-parametric methods in semi-exact control functionals \citep[SECF,][]{SECF}. More details about these methods are provided in Sections \ref{ssec:ZVCV}, \ref{ssec:CF} and \ref{ssec:SECF}, respectively. Theoretical properties of the methods are described, with further details available in Table 1 of \citet{South2022}. In addition to these choices of $\mathcal{H}$, choices based on neural networks have been considered \citep{Wan2020,Sun2023}.

All of the methods described in the subsections below are available for use in the ZVCV package on CRAN \citep{R_ZVCV}.

\subsection{Parametric bases} \label{ssec:ZVCV}
Consider a parametric basis $\mathcal{H} = \text{span}\{\phi_j\}_{j=1}^J$, where $J$ is the number of basis terms and $\phi_j: \mathcal{X} \rightarrow \mathbb{R}$ for $j=1,\ldots,J$. We can use this parametric basis to build control variates of the form $u(x) = \alpha + \sum_{j=1}^J \theta_j \mathcal{L}^{(2)} \phi_j(x)$, where $\alpha \in \mathbb{R}$ and $\theta \in \mathbb{R}^J$ are parameters to be optimised. This is the approach taken in ZVCV \citep{Assaraf1999,Mira2013}. Specifically, ZVCV uses the class of $q$th order polynomials, $\mathcal{H} = \text{span}\{x^{a}: a \in \mathbb{N}_0^d, \sum_{j=1}^d a_j \leq q\}$. A sufficient condition to check that ZVCV gives valid control variates for $\mathcal{X} = \mathbb{R}^d$ is that the tails of $\pi$ decay faster than polynomially.

To make this method more concrete, we will now provide forms for the control variates for first and second-order ZVCV. For first-order ZVCV, the polynomial is $h(x) = \theta_0 + \sum_{j=1}^d \theta_j x_j$ and the control variates are of the form
\begin{align*}
    u(x) &= \alpha + \mathcal{L}^{(2)} h(x)\\
    &= \alpha + \mathcal{L}^{(2)} \left[\theta_0 + \sum_{j=1}^d \theta_j x_j\right]\\
    &= \alpha + \Delta_x \left[\theta_0 + \sum_{j=1}^d \theta_j x_j\right] + \nabla_x \left[\theta_0 + \sum_{j=1}^d \theta_j x_j\right] \cdot \log \pi(x)\\
    &= \alpha + \sum_{j=1}^d \theta_j \nabla_{x_j} \log \pi(x) .
\end{align*}
This amounts to using the score function $\nabla_x \log \pi(x)$ as the control variates, or to using $\mathcal{U} = \text{span}\{1,s_1 ,\ldots,s_d\}$ where $s_j$ is shorthand for $\nabla_{x_j} \log \pi(x)$. Using a second-order polynomial with all two-way interactions, we get
\begin{align*}
\mathcal{U} &= \text{span}\{1,\tag{Constant corresponding to $\alpha$}\\
& \phantom{\text{span}\{1,} s_1,\ldots,s_d,\tag{Associated with first-order terms}\\
& \phantom{\text{span}\{1,}
2+2x_1s_1,\ldots,2+2x_ds_d,\tag{Associated with squared terms}\\
& \phantom{\text{span}\{1,}x_1s_2+x_2s_1,\ldots,x_{d-1}s_d + x_ds_{d-1}\}. \tag{Associated with two-way interaction terms}
\end{align*}
This can be extended to higher-order polynomials, where there will be ${{d+ q}\choose{d}}$ parameters to estimate.

Any of the approaches from Section \ref{ssec:optimisation} can be applied to estimate the parameters in ZVCV. Least squares is the approach most commonly associated with ZVCV, although its statistical performance tends to be poor when we do not have $n \gg J$ and its computational complexity is $\mathcal{O}({{d+q} \choose {d} }^3 + n{{d+q} \choose {d} }^2)$, assuming the polynomial includes all possible interactions. This cost quickly becomes unwieldy for high $d$ or $q$. Regularisation can be applied to help improve statistical performance. Alternatively, one can give some prior thought to what terms in the parametric basis are expected to be most helpful, as per the $\textit{a priori}$ technique in \citet{South2019}. For example, if the goal is to estimate the marginal mean for the second component, i.e. $f(x) = x_2$, then one may choose to replace a full $q$th order polynomial that would have ${{d+ q}\choose{d}}$ terms with $\mathcal{H} = \text{span}\{x_2,x_2^2,\ldots,x_2^q\}$.

ZVCV with a $q$th order polynomial is exact for $n>J$ when $\pi$ is Gaussian and $f$ is up to a $q$th order polynomial. This is referred to as the so-called ``semi-exactness" property. 
This property also indicates that the performance of ZVCV can be excellent when estimating means and variances in the big data limit. 

\begin{runningexample}

Continuing on with our bivariate Gaussian example, we can perform second-order ZVCV with least squares estimation using the following code.
\begin{Verbatim}[numbers=left,xleftmargin=5mm]
> CVs <- cbind(grads, 2+2*X*grads, X[,1]*grads[,2] + X[,2]*grads[,1])
> coef <- lm(f ~ CVs)$coef
> mean(f - CVs%*%coef[-1])
[1] -2.517344e-15
\end{Verbatim}
We already performed first-order ZVCV in Section \ref{ssec:example}.

\end{runningexample}

\subsection{Control functionals} \label{ssec:CF}

As an alternative to the parametric approach considered in ZVCV, \cite{Oates2017} consider a reproducing kernel Hilbert space for $\mathcal{H}$. 
We refer readers to \citet{Oates2017} for technical details about the reproducing kernel Hilbert space and we focus on the practical aspects of this approach. We also refer readers to \citet{Oates2017} and \citet{SECF} for (slightly) more interpretable forms for the boundary and tail conditions in this context.

Performing CF amounts to choosing a positive-definite kernel $k(x,y)$, for example a Gaussian or Mat\'ern kernel, and evaluating\footnote{The original CF paper \citep{Oates2017} has a $1+$ on the denominator, but this constant can also be selected to be zero, hence the simplified estimator. The original form also replaces $K_0$ with $\vect{K}_0 + \lambda n \vect{I}$ where $\lambda$ is a regularisation parameter, but alternative approaches for obtaining stable inverses are typically simpler to apply.}:
\begin{equation}\label{eqn:CF}
\hat{\mu}_{\text{CF}} =\frac{\vect{1}^T \vect{K}_0^{-1} f_n}{\vect{1}^T \vect{K}_0^{-1} \vect{1}}
\end{equation}
where $(f_n)_i = f(X^{(i)})$ and $(\vect{K}_0)_{i,j} = k_0(X^{(i)},X^{(j)})$. 
Using the first-order Langevin Stein operator, we have
\begin{align*}
k_0(x,y) &= \nabla_x^\top \nabla_y k(x,y) + (\nabla_x\log\pi)(x)^\top \nabla_y k(x,y)  \\
&\phantom{= }+ (\nabla_y\log\pi)(y)^\top \nabla_x k(x,y) + (\nabla_x\log\pi)(x)^\top (\nabla_y\log\pi)(y) k(x,y),
\end{align*}
whereas for a second-order Langevin Stein operator, we have
\begin{align*}
k_0(x,y) &= \mathcal{L}_x^{(2)}\mathcal{L}_y^{(2)} k(x,y)\\
&= \Delta_x \Delta_y k(x,y) + (\nabla_x\log\pi)(x)^\top \nabla_x\Delta_y k(x,y)  \\
&\phantom{= }+ (\nabla_y\log\pi)(y)^\top \nabla_y \Delta_x k(x,y) + (\nabla_x\log\pi)(x)^\top (\nabla_x \nabla_y^\top k)(x,y) (\nabla_y\log\pi)(y).
\end{align*}
\normalsize
Calculating $k_0$ involves gradients of the kernel so you can save time working out these gradients by using the explicit forms given in earlier papers, for example those given for radial kernels and second-order Langevin Stein operators in the supplementary material of \citet{SECF}.

Unlike previous methods, the optimisation behind the solution in \eqref{eqn:CF} is not based on least squares or on minimising the asymptotic variance. It is based on finding a minimum-norm interpolant of $u$ to $f$ \citep[see][for details]{Barp2018}. We discuss potential implications on the need (or lack of need) to remove burn-in in Section \ref{ssec:burnin}. To make inversion of $K_0$ possible, duplicate values that come from rejections in the MCMC chain must be removed. Numerical inversion of $K_0$ also typically requires numerical regularisation. The optimisation in CF does not optimise over the choice of kernel and its parameters, for example the length-scale parameter of the Gaussian kernel. One can instead pick these based on heuristics or using cross-validation.

The estimator in \eqref{eqn:CF} can be thought of as the posterior mean with a Gaussian process prior $f \sim \mathcal{GP}(0,K_0)$. The estimator looks different to typical control variate forms because of the interpolation of $u$ to $f$. 
However, if one separates out the samples for optimisation and evaluation of the integral, then one ends up with a form that is more similar to standard control variates. Interested readers, for example those applying the methods to unbiased MCMC or those wishing to perform cross-validation, may wish to refer to Equation 2 of \citet{Oates2017} for estimators that split the samples for optimisation and evaluation.

\citet{Oates2019} showed that CF can achieve better convergence rates than vanilla Monte Carlo, where the amount better depends on the smoothness of $f$, the smoothness of $\pi$ and the dimension. This property is not shared with ZVCV using a fixed order polynomial, for which the convergence rate remains the same but the constant can be improved. 
The computational complexity of CF is $\mathcal{O}(n^3 + n^2d)$, so it is standard to perform thinning prior to using the control variates. 
Despite this cubic dependence on $n$, \citet{Barp2018} proved in the MCMC context that CF is asymptotically more efficient than vanilla Monte Carlo for sufficiently smooth $f$ and $\pi$.

\begin{runningexample}

Continuing on with our Gaussian example, we will now perform CF with a first-order Stein operator. We use the Gaussian kernel $k(x,y) = \text{exp}(-\|x-y\|_2^2/\lambda^2)$ and following \citet{Garreau2017} we select
\begin{align*}
    \lambda^2=\frac{1}{2} \text{Med}\left\{ \|X^{(i)} - X^{(j)}\|^2: 1\leq i<j\leq n\right\},
\end{align*}
where $\text{Med}$ is the empirical median. Code for performing this with the ZVCV package is below:
\begin{Verbatim}[numbers=left,xleftmargin=5mm]
> library(ZVCV)
> lambda <- medianTune(unique(X))
> CF(f,X,grads,steinOrder=1,kernel_function="gaussian",sigma=lambda)
$expectation
           [,1]
[1,] 0.06374077
\end{Verbatim}

Alternatively, we can perform this from scratch. We first remove duplicate samples (Lines 1-6) and then we calculate the matrix of squared norms for the median heuristic and kernel calculations (Lines 8-11). Line 9 calculates the squared norm and stores it in an efficient way (storing the lower-triangular elements only). Line 10 calculates the median heuristic and Line 11 converts the squared norms into a matrix. We then calculate the $K_0$ matrix, where this code is by no means the most efficient code possible. The $K_0$ matrix typically requires numerical regularisation, which is what the function \verb|nearPD| does. The final estimate is reported on Line 37. 
\begin{Verbatim}[numbers=left,xleftmargin=5mm]
> # Removing duplicates
> dups <- which(duplicated(X))
> Xu <- X[-dups,]
> gradsu <- grads[-dups,]
> fu <- f[-dups]
> nu <- nrow(Xu)
> 
> # Calculating matrix of squared norms
> Z <- dist(Xu)^2
> lambda2 <- 0.5*median(Z)
> Z <- as.matrix(Z)
> 
> # Calculating the K0 matrix
> d <- 2
> k <- exp(-Z/lambda2)
> K0 <- matrix(NaN,nrow=nu,ncol=nu)
> for (i in 1:nu){
+   for (j in i:nu){
+     x <- Xu[i,]
+     y <- Xu[j,]
+     
+     k_ <- k[i,j]
+     k_gradx <- -2/lambda2*k_*(x-y)
+     k_grady <- 2/lambda2*k_*(x-y)
+     k_gradxgrady <- -4/lambda2^2*k_*Z[i,j] + 2/lambda2*k_*d
+     
+     K0[i,j] <- K0[j,i] <- k_gradxgrady + t(gradsu[i,])%*%k_grady + 
+       t(gradsu[j,])%*%k_gradx + t(gradsu[i,])%*%gradsu[j,]*k_
+   }
+ }
> 
> # Getting the estimate
> K0 <- nearPD(K0) # For numerical stability in the inverse
> K0inv <- solve(K0)
> t(rep(1,nu))%*%K0inv%*%fu/(t(rep(1,nu))%*%K0inv%*%rep(1,nu))
           [,1]
[1,] 0.06374077
\end{Verbatim}

The estimate from this single run does not appear better than vanilla Monte Carlo, but the variance across 100 runs is about 8-9 times lower with CF than with vanilla Monte Carlo. This could also be improved by picking $\lambda$ with cross-validation.

\end{runningexample}

\subsection{Semi-exact control functionals} \label{ssec:SECF}

In SECF, the parametric and non-parametric approaches of ZVCV and CF are combined to achieve the improved convergence rates of CF and the semi-exactness of ZVCV.

The control variate takes the form
\begin{equation*}
    u(x) = \alpha_0 + \underbrace{\sum_{j=1}^{J-1} \alpha_{j} \mathcal{L}^{(2)} \phi_j (x)}_{\text{related to ZV-CV}} + \underbrace{\sum_{i=1}^n \theta_{i} k_0(x, X^{(i)})}_{\text{related to CF}}.
\end{equation*}
where $\{\phi_j\}_{j=1}^{J-1}$ is a parametric basis and the aim is to optimise $\alpha \in \mathbb{R}^J$ and $\theta \in \mathbb{R}^n$.

The optimisation of $\alpha$ and $\theta$ is performed to achieve a minimum-norm interpolant with the constraint that the estimator must be semi-exact, i.e. we must have $u = f$ whenever $f \in \mathcal{F} = \text{span}(1,\mathcal{L}^{(2)}\phi_1,\ldots,\mathcal{L}^{(2)}\phi_{J-1})$. This amounts to solving the linear system
\begin{equation*}
    \begin{bmatrix} \vect{K}_0 & {\Phi} \\ {\Phi}^T & \vect{0} \end{bmatrix} \begin{bmatrix} \vect{\theta} \\ \vect{\alpha} \end{bmatrix} = \begin{bmatrix} f_n \\ \vect{0} \end{bmatrix}, \text{ where } \vect{\Phi} = \begin{bmatrix} 1 & \mathcal{L}^{(2)} \phi_1(X^{(1)}) & \cdots & \mathcal{L}^{(2)} \phi_{J-1}(X^{(1)}) \\ \vdots & \vdots & \ddots & \vdots \\ 1 & \mathcal{L}^{(2)} \phi_1 (X^{(n)}) & \cdots & \mathcal{L}^{(2)} \phi_{J-1}( X^{(n)}) \end{bmatrix} ,
\end{equation*}
for which the closed-form solution is
\begin{align*}
\hat{\mu}_{\text{SECF}} &= \hat{\alpha}_0\\
&= \vect{e}_1^T ( \vect{\Phi}^T \vect{K}_0^{-1} \vect{\Phi} )^{-1} \vect{\Phi}^T \vect{K}_0^{-1} f_n,
\end{align*}
where $e_1 = (1,0,\ldots,0)^\top$. This can also be thought of as using a Gaussian process prior $f \sim \mathcal{GP}(\sum_{j=1}^{J-1}\alpha_j \mathcal{L}\phi_j ,{K}_0)$, so the difference to CF is that we have informed the prior mean using the parametric approach.

SECF can outperform both ZVCV and CF for sufficiently large $n$ but its computational complexity is $\mathcal{O}(N^3 + J^3)$, or $\mathcal{O}(N^3 + {{d+q} \choose {d}}^3)$ if the parametric basis is a $q$th order polynomial. This can be sped up by thinning, approximate kernel methods like Nystr\"{o}m approximations and using approximate solutions to the linear system.

\begin{runningexample}

We will now run SECF on the Gaussian example. We will use a first-order polynomial for the parametric basis and a Gaussian kernel as in the CF implementation for the non-parametric component. Code for performing this with the ZVCV package is below:
\begin{Verbatim}[numbers=left,xleftmargin=5mm]
> SECF(f,X,grads,polyorder=1,steinOrder=1,kernel_function="gaussian",sigma=lambda)
$expectation
              [,1]
[1,] -4.440892e-15
\end{Verbatim}

Alternatively, performing this from first principles we obtain the following.
\begin{Verbatim}[numbers=left,xleftmargin=5mm]
> Phi <- cbind(1,gradsu)
> solve(t(Phi)%*%K0inv%*%Phi,t(Phi)%*%K0inv%*%fu)[1]
[1] 2.727166e-13
\end{Verbatim}

\end{runningexample}

\section{Bespoke control variates for MH-MCMC} \label{sec:MH}

This section describes bespoke control variates that have been developed in \citet{Hammer2008} and \citet{Delmas2009} for MH samplers. Extensions to reversible-jump MCMC and a mode-jumping algorithm can be found in \citet{Hammer2008}. Alternative control variates for MH-MCMC with independent proposals can be found in \citet{Atchade2005}.

These methods require storage of the chain $X^{(i)}$ for $i=1,\ldots,n$, the proposals $X^{(i+1)^*} \sim q(\cdot|X^{(i)})$ for $i=1,\ldots,n-1$, the MH ratios $R(X^{(i)},X^{(i+1)^*})$ for $i=1,\ldots,n-1$ and potentially the acceptance decisions $\delta_i\in \{0,1\}$ indicating whether $X^{(i+1)^*}$ was accepted ($\delta_i=1$) for $i=1,\ldots,n-1$. The control variates then work on an augmented space including the proposals and accept decisions.
The variance reduction achieved through these control variates is typically fairly modest, with \citet{Hammer2008} reporting variance reductions of 15\% to 35\%. 

\citet{Hammer2008} propose to use estimators of the form
\small
\begin{align*}
\hat{\mu}_{\text{CV}} = \frac{1}{n-1} \sum_{i=1}^{n-1} \left[f(X^{(i+1)}) - u(X^{(i)},X^{(i+1)^*},\delta_i)\right] + \underbrace{\int_{\mathcal{X}} \int_{\mathcal{X}} \left[\alpha(x,y)u(x,y,1) + (1-\alpha(x,y))u(x,y,0)\right] q(y|x) \pi(x) \mathrm{d} x\mathrm{d} y}_{\dagger},
\end{align*}
\normalsize
where the expectation of $u$ on the augmented space, $\dagger$, is zero. The authors do this by selecting
\begin{align*}
    u(X^{(i)},X^{(i+1)^*},\delta_i) = w_1(X^{(i)},X^{(i+1)^*},\delta_i) f(X^{(i)}) + w_2(X^{(i)},X^{(i+1)^*},\delta_i)f(X^{(i+1)^*})
\end{align*}
with carefully chosen $w_1$ and $w_2$.

A simple choice is $w_1(x,y,\delta) = - h(x,y) \frac{R(x,y)}{1+R(x,y)}$ and $w_2(x,y,\delta) =  h(y,x) \frac{R(x,y)}{1+R(x,y)}$, where $h(x,y)$ is a function to be specified. A more sophisticated choice including the accept decision is to use $w_1(x,y,\delta) = -  \delta h(y,x)(1-\alpha(y,x))$ and $w_2(x,y,\delta) =  (1-\delta)h(x,y)\alpha(x,y)$. There is flexibility in the choice of $h$, with a good default being to use $h(x,y) = \theta$ where $\theta$ is a constant to be optimised. With this choice, the simple control variate reduces to $u(x,y,\delta) = \theta \frac{R(x,y)}{1+R(x,y)} (f(y) - f(x))$, i.e.\ it is proportional to the difference in function evaluations multiplied by the acceptance probability of \citet{Barker1965}. As with previous control variates, $\theta$ needs optimising and one could also introduce an additive constant $\alpha$ to be optimised.  The authors choose to select $\theta$ to minimise a batch means estimate of the asymptotic variance.

\citet{Delmas2009} propose an alternative form for the control variates based on a reformulation of waste-free Monte Carlo \citep{Ceperley1977}. Their control variates are of the form 
\begin{align*}
u(X^{(i)},X^{(i+1)^*},X^{(i+1)}) &= \mathbb{E}[h(X^{(i+1)})|X^{(i)},X^{(i+1)^*}] - h(X^{(i+1)})\\
&= \alpha(X^{(i)},X^{(i+1)^*})h(X^{(i+1)^*}) + (1-\alpha(X^{(i)},X^{(i+1)^*}))h(X^{(i)}) - h(X^{(i+1)}).
\end{align*}
The authors found that the asymptotic variance of the estimator with these control variates is typically larger than the vanilla Monte Carlo estimator, although they did not include a multiplicative factor or an additive constant to help optimise the variance reduction. Control variates of the form $\alpha + \theta  \mathbb{E}[h(X^{(i+1)})|X^{(i)},X^{(i+1)^*}] - \theta h(X^{(i+1)})$, or including a multiplicative factor $\theta$ in $h(x)$ itself, should offer better performance in practice.

\begin{runningexample}

Let us finish our Gaussian example with the control variates of \citet{Hammer2008} and \citet{Delmas2009}. These control variates require storage of the samples, proposals and MH ratios. 

We will refer to \verb|X|, which is the $n \times 2$ matrix storing $X^{(1)},X^{(2)},\ldots,X^{(n)}$, and to \verb|Xprop|, which is the $(n-1) \times 2$ matrix storing $X^{(2)^*},\ldots,X^{(n)^*}$. We will also refer to the ($n-1$)-vector \verb|R| containing the MH ratios and the ($n-1$)-vector \verb|accept| containing the accept decisions.

The code for performing these methods is provided below. Again, we perform least squares for simplicity but we note that alternative estimators for the coefficients are possible. The estimators appear to be performing worse than vanilla Monte Carlo based on this single run. Their performance is similar to vanilla Monte Carlo over 100 runs, with about a 5\% improvement using \citet{Hammer2008} and a 2\% deterioration using \citet{Delmas2009}. This may be in part due to the high acceptance rate of the sampler, leaving little to be gained from the proposals.
\begin{Verbatim}[numbers=left,xleftmargin=5mm]
> # Calculating the integrand of interest
> f_curr <- X[-n,1]
> f_prop <- Xprop[,1]
> f_next <- X[-1,1]
> 
> # Calculating the MH acceptance probability from the ratios
> MH <- pmin(1,R)
> MH_reverse <- pmin(1,1/R)
> 
> # Hammer and Tjelmeland 
> CVs <- (1-accept)*MH*f_prop - accept*(1-MH_reverse)*f_curr
> beta <- lm(f_curr~CVs)$coefficient
> mean(f_curr - CVs*beta[-1])
[1] -0.04693029
> 
> # Delmas and Jourdain
> CVs <- MH*f_prop + (1-MH)*f_curr - f_next
> beta <- lm(f_curr~CVs)$coefficient
> mean(f_curr - CVs*beta[-1])
[1] -0.04261975
\end{Verbatim}

\end{runningexample}

\section{Further practical advice} \label{sec:practical}

This section provides further practical advice and an example to assist readers in making decisions about the use of control variates. Control variates for MCMC are still a relatively new development compared to other areas of MCMC, especially when it comes to theory. The majority of this advice is based on empirical evidence.

\subsection{Should I use control variates for my problem?}

The purpose of control variates is to reduce the variance in a more efficient way than running the MCMC chain for longer. If we are already sufficiently happy with the quality of our estimator then using control variates would be unnecessary. If not, then we should consider the cost of increasing $n$ versus the costs of using control variates. Increasing $n$ means running the MCMC chain for more iterations and evaluating $f$, while using control variates means selecting $u \in \mathcal{U}$ and evaluating $u$. For this reason, control variates are most useful when increasing $n$ is expensive.

Importantly, the trade-off also depends on the variance reduction that would be achieved with control variates. For example, if control variates reduced the variance by a factor of 2.5 then we would want the cost of applying control variates to be less than the cost of adding another $1.5n$ iterations. We generally do not know exactly what variance reductions we will achieve, but we might have a good idea of whether the control variates will perform well.

\subsection{Which control variate should I use?}

The majority of this chapter has been dedicated to providing families of control variates for MCMC so how do we know which one we should use? This is a challenging question and the answer is not always straightforward. 

Your choice should be partially dictated by the type of sampler you are using. When using a gradient-based sampler, the approaches from Section \ref{sec:GBCV} are a natural selection. 
If gradients are not already available, then the computational cost of computing them for the control variates in Section \ref{sec:GBCV} may outweigh the potential benefits. In the context of applications where random-scan Gibbs sampling would be feasible (even if you are not using it), you could consider using the control variates $h - Ph$ with $P$ based on random-scan Gibbs as in Section \ref{sec:PoissonCV}. If you are using MH MCMC, then there are applications where the control variates presented in Sections \ref{sec:PoissonCV}-\ref{sec:MH} will all be available to you.

You should also consider the complexity and smoothness of $\pi$ and $f$.  If $\pi$ is close to Gaussian and the goal is to estimate the $q$th moment of $\pi$, then ZVCV with a $q$th order polynomial or $h-Ph$ using $P$ from random-scan Gibbs sampling and $h$ with a $q$th order polynomial are good choices. When $f$ and $\pi$ are smooth and $d$ is reasonably low-dimensional, CF is the natural choice. When interest is in non-smooth $f$, for example an indicator function for estimating a probability, then the gradient-based approaches of Section \ref{sec:GBCV} are unlikely to be helpful.

The dimension of your problem is also a factor. When the dimension is high relative to $n$, the kernel methods of CF and SECF are unlikely to offer much variance reduction. You may wish to restrict yourself to parametric approaches based on a subset of the dimensions of interest. These approaches will keep the number of parameters to estimate manageable.

Finally, you should consider the cost of using control variates in terms of storage and computation time. Gradient-based control variates require computation and storage of (at least some) gradients, while the control variates in Section \ref{sec:MH} require storage of proposals alongside the MCMC chain. This is not necessarily true of control variates based on the Poisson equation. The optimisation step for some methods, for example first-order ZVCV, is naturally relatively inexpensive while the optimisation for more complex control variates, for example SECF, can be very costly. If it is particularly expensive to gain more MCMC samples, you may choose to focus almost entirely on statistical performance regardless of the computational cost in performing control variates. Otherwise, you may wish to focus on the cheaper control variates or on tricks like the use of stochastic optimisation \citep{Si2020}. 

A promising but slow approach to choosing a control variate family is to perform cross-validation. 
This may involve estimating $u$ on a subset of the samples and evaluating a performance measure such as the MSE between $f$ and $u$ on the remaining samples. However, this approach adds substantially to the computational cost and also requires splitting samples for optimisation and evaluation.

\subsection{Multiple functions of interest}

The control variates described here are based on estimating a single function $f$ at a time. Suppose there are $r$ functions of interest, for example when estimating all marginal means and variances we would have $r=2d$. The approaches described above call for $r$ separate optimisation procedures to estimate the optimal control variates for each function. This is computationally convenient for some optimisation methods, for example least squares, but none of the methods exploit the possible correlation between the $r$ functions to improve the coefficient estimates. \citet{Sun2023} use ideas from transfer learning to estimate control variates for multiple functions simultaneously. Their approach is an extension to CF, but one could also consider extending parametric approaches by borrowing ideas from multi-task regression.

\subsection{Burn-in and bias removal}\label{ssec:burnin}

Perhaps unexpectedly, several of the estimators considered in this chapter are capable of correcting bias in certain situations.

The non-parametric Stein-based control variates (CF and SECF) can offer bias correction in the sense that they can provide consistent estimators when the Markov chain is not $\pi$-invariant in some circumstances \citep[see Theorem 1 of][for example]{SECF}. This raises an interesting question of whether one is better off running a slow $\pi$-invariant Markov chain or a fast but biased MCMC algorithm that is corrected for in post-processing. Burn-in removal should not be necessary for these methods. 

It seems that little is known about the bias-correcting properties of estimators based on Poisson equations. \citet{Dellaportas2012} provide some initial comments on the potential for removal of bias from an initial burn-in period, but they still recommend burn-in removal.

For other methods, for example ZVCV, it would be sensible to use standard approaches to remove burn-in (see Chapter 4). 

\subsection{Thinning}

Thinning is known to increase the variance of estimators \citep{Geyer1992}, so it is generally ill-advised from a statistical perspective. It can also lead to worse estimators of control variate coefficients, giving another reason to avoid thinning. For these reasons, it is not recommended for the methods that have computational complexity $\mathcal{O}(n)$, which includes ZVCV and the majority of methods from Sections \ref{sec:PoissonCV} and \ref{sec:MH}. However, CF and SECF both have computational complexity $\mathcal{O}(n^3)$, making them costly for high $n$. For large $n$, for example $n>1000$, these methods become prohibitively slow. The $\mathcal{O}(n^2)$ storage requirements also become an issue for these methods when $n$ is large. We therefore recommend thinning for CF and SECF when $n>1000$. 

\subsection{Parameterisation}

The performance of control variates is dependent on the parameterisation, meaning that one could consider applying a deterministic, invertible transformation and working on this transformed space for post-processing. This is particularly true of Stein-based control variates, where one could consider applying a transformation $\tilde{x} = \psi(x)$, replacing $x$ with $\tilde{x}$, $f(x)$ with $f(\psi^{-1}(\tilde{x}))$ and $\nabla_{x} \log \pi(x)$ with $\nabla_{\tilde{x}} \left(\log \pi(\psi^{-1}(\tilde{x})) + \log(| \frac{d\psi^{-1}(\tilde{x})}{d\tilde{x}}|)\right)$ in all notation from Section \ref{sec:GBCV}. In our experience, common transformations used to shift parameters to the real line work well for Stein-based control variates too. They tend to offer higher variance reductions and are more likely to meet the required regularity conditions than parameters on constrained spaces

\subsection{Application to Bayesian variable selection}\label{ssec:BVS}

We conclude with an application of control variates to Bayesian variable selection (BVS) for linear regression models. 
In BVS, a vector of binary variables $\gamma = (\gamma_1,..., \gamma_p)\in\{0,1\}^p$ is used to index which variable is included or ``active'' in the model ($\gamma_i=1$ indicates the $i$th variable is included and it is excluded if $\gamma_i=0$). The vector $\gamma$ is given a prior inclusion probability of $p(\gamma_i =1) = 1/p$ for $i=1,...,p$. We follow the BVS framework of \cite{Chipman} which considers the collection of models
\begin{align*}
    {y} | \beta_{\gamma}, \gamma, \sigma^2 &\sim \mathcal{N}({X}_{\gamma}\beta_{\gamma}, \sigma^2I_n),\\
    \beta_{\gamma}|\gamma, \sigma^2 &\sim \mathcal{N}(0, 10^3(X_{\gamma}^TX_{\gamma})^{-1}\sigma^2)\\
    p(\sigma^2) &\propto \sigma^{-2}    
\end{align*}
where ${y}$ is the response vector of length $n$, ${X}_{\gamma}$ is an $n\times |\gamma|$ submatrix of the $n\times p$ predictor matrix corresponding to the active columns of ${X}$, and $\beta_{\gamma}$ is the vector of regression parameters of length $|\gamma|$. We consider a simulated dataset similar to Example 5.1 of \cite{Zanella} where the first two covariates are highly correlated and have good explanatory power. We use $n=70$ and $p=5$.

For this model, it is possible to marginalise over the parameters and variance so that sampling may take place directly over the space $\{0,1\}^p$. Gibbs samplers are the typical approach to BVS and proceed by updating components of $\gamma$ via their full conditional probability $\gamma_i \sim \pi(\gamma_i | {y}, \gamma_{-i})$. We use a random scan Gibbs sampler which selects this index $i\in\{1,...,p\}$ at random and applies this update. We refer the reader to Supplementary Section B.1 of \citet{Zanella} for an explicit, computationally efficient expression of $\pi(\gamma_i | {y}, \gamma_{-i})$.

\begin{figure}[h]
    \centering
    \includegraphics[width=0.7\linewidth]{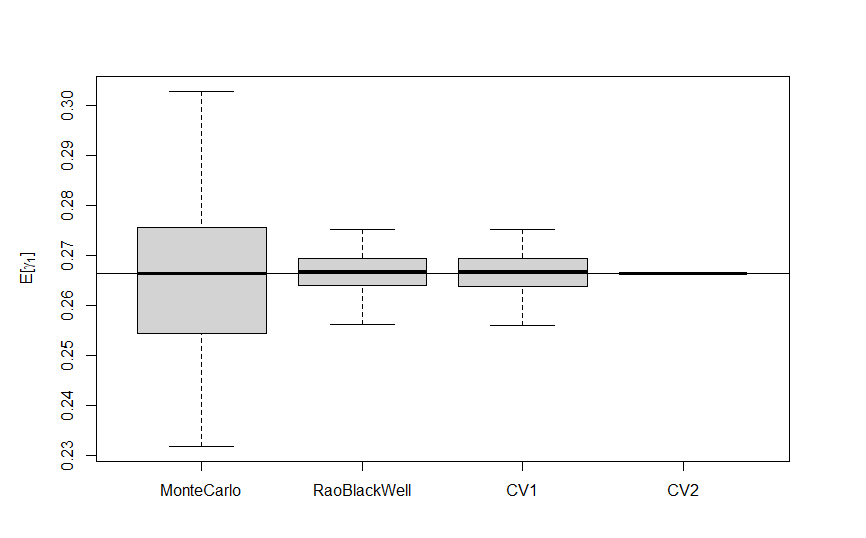}
    \caption{Boxplots of estimates of $\mathbb{E}_{\pi}[\gamma_1]$ from 100 independent MCMC runs with $n=10,000$. The gold standard line was calculated from a single long run of the Gibbs sampler.}
    \label{fig:bvs1}
\end{figure}

\begin{figure}[h]
    \centering
    \includegraphics[width=0.7\linewidth]{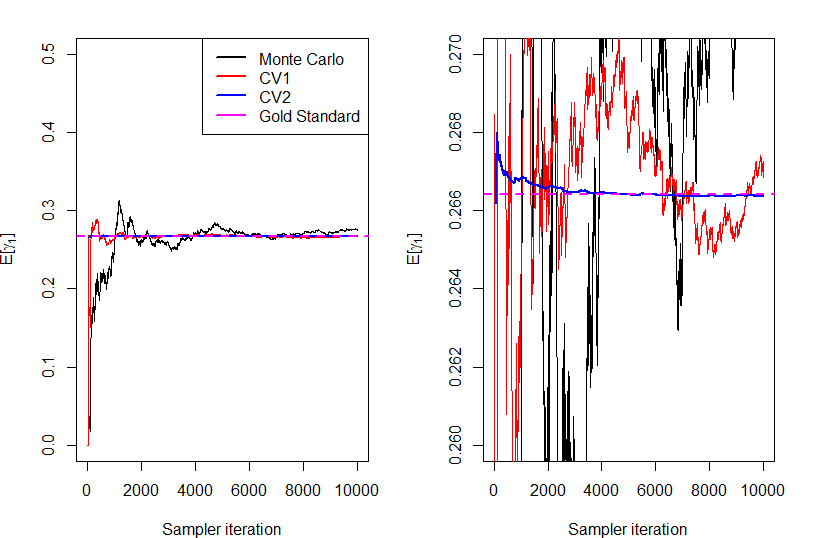}
    \caption{Convergence of the estimators of $\mathbb{E}_{\pi}[\gamma_1]$. Results are shown for iteration three onwards. The control variate estimates are based on re-estimating the coefficients for each new iteration, so the performance for low iterations is actually indicative of what performance would be like for low $n$. A zoomed-in version of the convergence is shown on the right. }
    \label{fig:bvs2}
\end{figure}

Our goal is to estimate the inclusion probability for the first variable, i.e.\ $\mathbb{E}_{\pi}[\gamma_1]$. We cannot apply gradient-based control variates here because the space is discrete and does not admit gradients. Instead, we will construct control variates based on the description in Section \ref{sec:PoissonCV} with the function $h(\gamma) = \theta_1\gamma_1 + \theta_2\gamma_2$. The one-step-ahead conditional expectation operator can be computed explicitly as,
$$
P \gamma_i = \frac{1}{p}\pi(\gamma_i = 1 | \gamma_{-i}, {y}) + \frac{p-1}{p}\gamma_i,
$$
for $i=1,2$. This gives rise to control variates of the form,
\begin{align*}
u(\gamma) &= \alpha + h(\gamma) - Ph(\gamma) \\
&= \alpha + \sum_{i=1}^2\theta_i\left[  \gamma_i - \frac{1}{p}\pi(\gamma_i = 1 | \gamma_{-i}, {y}) - \frac{(p-1)}{p}\gamma_i\right].
\end{align*}
We will estimate the coefficients using least squares because of its semi-exactness properties. Interestingly, for $\theta_1 = p$, and $\theta_2 = 0$ the control variate estimator corresponds to the Rao-Blackwellised estimator $\frac{1}{n}\sum_{i=1}^n\mathbb{E}_{\gamma_1 \sim \pi(\cdot|\gamma_{-1},{y})}[\gamma_1]$ \cite[Section B.3]{Zanella}. We therefore consider the full control variates (denoted CV2 in figures), as well as the special case with $\theta_2=0$ (denoted CV1). The control variate estimator with $\theta_2=0$ will select $\theta_1 \approx p$ using least squares to obtain similar results to the Rao-Blackwellised estimator. As shown in Figures \ref{fig:bvs1} and Figure \ref{fig:bvs2}, the control variate CV2, which uses information from both correlated variables $\gamma_1$ and $\gamma_2$, gives much better performance. The code for applying control variates to a single MCMC run is shown below, with extended code available at \url{https://github.com/LeahPrice/CVBookChapter/}. 

\begin{Verbatim}[numbers=left,xleftmargin=5mm]
> # Standard Monte Carlo estimator
> f <- gamma_1
> mean(f)
[1] 0.275
> 
> # Using a single function h(gamma) = gamma_1
> CV1 <- gamma_1 - (1/p*p_gamma_1 + (p-1)/p*gamma_1)
> coefs1 <- lm(f ~ CV1)$coef
> mean(f - coefs1[-1]*CV1)
[1] 0.2668885
> 
> # Using h(gamma) = theta_1*gamma_1 + theta_2*gamma_2
> CV2 = cbind(CV1, gamma_2 - (1/p*p_gamma_2 + (p-1)/p*gamma_2))
> coefs2 <- lm(f ~ CV2)$coef
> mean(f - CV2%*%coefs2[-1])
[1] 0.2663682
\end{Verbatim}

This example highlights the utility of control variates for challenging applications. Rao-Blackwellisation is a popular approach for improving estimates in BVS and we show that control variates can offer substantial improvements on this. More broadly, the numerous approaches provided for constructing control variates in this chapter have the potential to improve estimates of expectations in a broad range of practical applications.

\section*{Acknowledgements}
This work was supported by the Australian Research Council under grants DE240101190 and DE250101447.

\bibliographystyle{apalike} 
\bibliography{handbook}

\end{document}